\documentclass[aps,prl,float,twocolumn,showpacs,preprintnumbers,amsmath,amssymb,groupedaddress]{revtex4-1}

\usepackage{graphicx}  
\usepackage{dcolumn}   
\usepackage{bm}        
\usepackage{color}     
\usepackage{rotating}  
\newcommand{\llangle}{\langle\hspace{-0.5mm}\langle}
\newcommand{\rrangle}{\rangle\hspace{-0.5mm}\rangle}

\begin{document}
\title{Spin-orbit coupling in hydrogenated graphene}
\author{Martin Gmitra, Denis Kochan, and Jaroslav Fabian}
\affiliation{Institute for Theoretical Physics, University of Regensburg, 93040 Regensburg, Germany\\
}




\begin{abstract}
First-principles calculations of the spin-orbit coupling in graphene with hydrogen adatoms in dense and dilute limits are presented.
The chemisorbed hydrogen induces a giant local enhancement of spin-orbit coupling due to $sp^3$ hybridization which depends
strongly on the local lattice distortion. Guided by the reduced symmetry and the local structure of the induced dipole moments
we use group theory to propose realistic minimal Hamiltonians that reproduce the relevant spin-orbit effects for both
single-side semihydrogenated graphene (graphone) and for a single hydrogen adatom in a large supercell. The principal linear spin-orbit band
splittings are driven by the breaking of the local pseudospin inversion symmetry and the
emergence of spin flips on the same sublattice.
\end{abstract}

\maketitle

Spin-orbit coupling is central for a variety of spintronics phenomena
\cite{Zutic2004:RMP, Fabian2007:APS} such as
spin relaxation, spin transport, or topological
quantum spin Hall effects. Itinerant electrons in graphene have weak spin-orbit
coupling as they are formed from $p_z$ orbitals. The Dirac cones are separated
by what is called intrinsic spin-orbit coupling
of $2\lambda_{\rm I}=24\,\rm{\mu eV}$ due to $p_z$-$d$ mixing
\cite{Gmitra2009:PRB, Konschuh2010:PRB, Konschuh2012:PRB, Abdelouahed2010:PRB}.
This small value is desirable for long spin lifetimes, but experiments
suggest \cite{Tombros2007:N, Pi2010:PRL, Yang2011:PRL, Mani2012:NC}
that spin relaxation is governed by much stronger spin-orbit,
or perhaps magnetic \cite{Lundeberg2012},
interactions than the intrinsic one. Potential culprits are light adatoms
\cite{Neto2009:PRL, Ertler2009:PRB, Zhang2012:NJP} which are
typically not important for momentum scattering but may be essential for spin-flip
scattering. On the other hand, large spin-orbit coupling, when controlled,
is desirable for engineering robust quantum topological phases in graphene
covered with heavy adatoms \cite{Qiao2010:PRB, Weeks2011:PRX, Zhang2012:PRL}.

Hydrogen is an ideal light adatom to study induced spin effects in graphene.
Not only can it produce local magnetic moments \cite{Yazyev2008:PRL,Fiori2010:PRB, Sofo2012:PRB, Soriano2011:PRL},
as recently experimentally demonstrated \cite{McCreary2012:PRL}, but
it should also enhance graphene's spin orbit coupling (SOC), as proposed in
Ref. \onlinecite{Neto2009:PRL}. Unlike for heavy adatoms whose cores
can directly contribute to SOC, the enhancement of SOC from hydrogen
is solely due to $sp^3$ hybridization facilitated by local structural deformation.
The SOC effects can thus directly probe $sp^3$ phenomena in graphene.

The presence of both magnetic moment and large spin-orbit coupling makes
the spin physics exciting, but also challenging to explain the spin relaxation experiments.
To disentangle the two contributions, as well as to see what new phenomena they
can lead to, it is important to consider them separately. Here we present
a quantitative and qualitative study of SOC induced by hydrogen on graphene
in two limits. One is the dense limit, represented here by single-side
semihydrogenated graphene (also called graphone) \cite{Zhou2009:NanoLett}.
This structure is relatively simple and allows for
a quantitative analysis of the $sp^3$ hybridization effects on various spin-orbit
parameters. Our results indeed show a giant enhancement of SOC, strongly
dependent on the buckling deformations of this structure. We introduce a
single-band and tight-binding Hamiltonians to describe the main SOC effects.
In particular, we show how pseudospin inversion asymmetry (PIA) introduces new
terms (we call them PIA SOC), which couple the opposite spins on the
same sublattice, in addition to Bychkov-Rashba nearest neighbor hoppings.

We also quantify the local spin-orbit structure and
propose a minimal realistic SOC Hamiltonian in the dilute limit, represented
by large supercells (starting with $5 \times 5$), intensively studied for orbital
effects \cite{Duplock2004:PRL,Pereira2006:PRL,Robinson2008:PRL,Wehling2010:PRL}.
Based on first-principles calculations we demonstrate a giant local---and we identify 
the local impurity region from the dipole moments distribution---enhancement 
of SOC due to $sp^3$ hybridization and formulate a minimal realistic SOC hopping model. 
We believe that this is a benchmark model to study spin relaxation, spin transport, 
but also weak (anti)localization \cite{McCann2012:PRL, Lundeberg2012} phenomena 
in which spin-flip and spin-orbit scattering play an important role.

\paragraph{Dense limit: single-side semihydrogenated graphene (SHG).}

\begin{figure}[h!]
\centering
\includegraphics[width=0.99\columnwidth,angle=0]{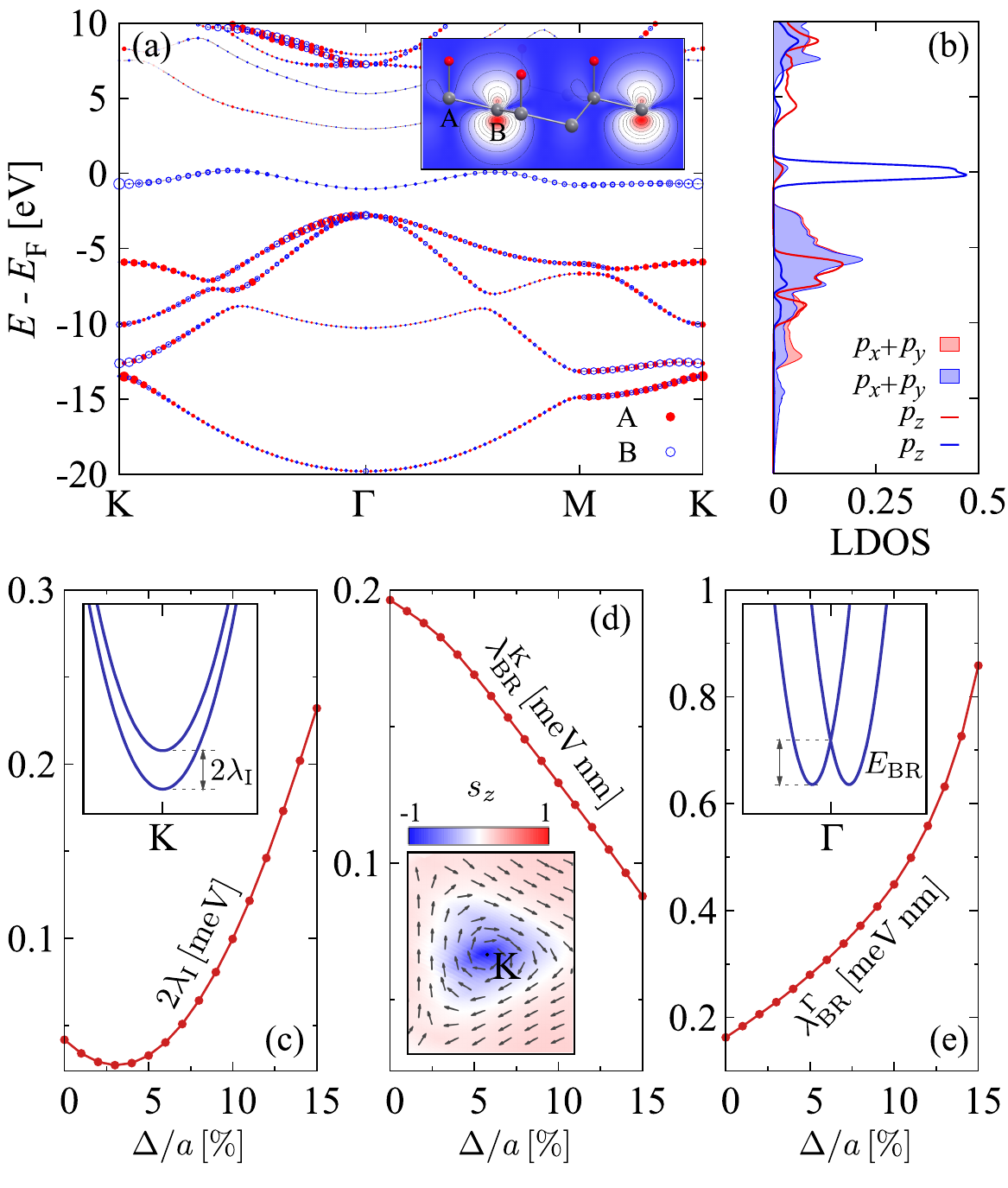}
\caption{(Color online) Top: Calculated electronic band structure of
single-side semihydrogenated graphene.
(a)~Sublattice resolved band structure for the distortion $\Delta / a=14 \%$.
The filled (red) circles correspond to sublattice A whereas the
open (blue) to sublattice B. The circles radii correspond to the carbon atom charge densities.
The inset shows the structure and the probability density of the
flat band at the $\rm K$ point.
(b)~Orbital resolved local density of states. Bottom:
Extracted spin-orbit coupling parameters
for the $\pi^*$ band at $\rm K$ and $\Gamma$
as functions of $\Delta/a$:
(c)~Intrinsic spin-orbit coupling splitting $2\lambda_{\mathrm{I}}$ at $\rm K$. The inset
shows the band splitting.
(d)~Adatoms-induced SOC splitting $\lambda^{\rm{K}}_{\rm BR}$ at $\rm K$.
The inset shows the spin texture around $\rm K$ for the
lower spin-orbit split band. The in-plane components are shown by the arrows
while the $z$ component by the color map.
(e)~Adatoms-induced SOC splitting $\lambda^{\Gamma}_{\rm BR}$ at $\Gamma$. The inset shows
the band splitting around $\Gamma$ with the identified
Bychkov-Rashba energy $E_{\rm BR}=0.87\,{\rm \mu eV}$ for $\Delta/a = 14\%$.
}
\label{Fig:SHG}
\end{figure}

The effects of $sp^3$ hybridization on SOC are studied using
single-side semihydrogenated graphene [inset in Fig.~\ref{Fig:SHG}(a)]
with different degrees of out-of-plane lattice distortion $\Delta$ of the
hydrogenated carbon site which is on sublattice A. The C-H bond length 
$d_{\rm H}$ is $1.13$~$\rm\AA$, and we take the lattice constant to be the 
relaxed distance $a \approx 2.516 $ $\rm\AA$ between the nearest neighbors
around the adatom site in the supercell calculations below;  
in graphene the lattice constant is $2.46$ $\rm\AA$. (The relaxed SHG structure of 
lattice constant $2.535~{\rm\AA}$ would have $\Delta/a = 9.7\%$ and the 
$sp^3$ tetrahedron 20.41\%.) In examples we choose $\Delta/a=14\%$ which corresponds
to relaxed large supercell structures.

To study SOC effects we restrict the computational basis to be spin unpolarized.
The calculated electronic structure and the projected local
density of states for SHG are shown in Fig.~\ref{Fig:SHG}(top) for $\Delta/a=14\%$.
Compared to graphene, in which $\pi$ and $\pi^*$ bands without SOC touch
at K \cite{Wallace1947:PR}, the C-H bonding pulls them apart: the $\pi$ band, which at $\rm{K}$
comes mainly from sublattice $\rm{A}$, is shifted to about 5~eV below the Fermi level
(the GW approximation predicts a greater shift by about 2 eV \cite{Fiori2010:PRB}).
The $\pi^*$ band, which comes from sublattice $\rm{B}$, lies at the Fermi level. This
band, which we consider for our SOC analysis, is narrow since
the nearest-neighbor hopping is inhibited for $p_z$ electrons on $\rm{B}$.
The inset in Fig.~\ref{Fig:SHG}(a) shows the $\pi^*$ probability density at K that
has the $p_z$ character on sublattice B.

We now extract the SOC parameters for the states at K and $\Gamma$ for the $\pi^*$ band which is at the Fermi level.
Since this band is nondegenerate, the effective spin-orbit Hamiltonian can be expressed
via the spin Pauli matrices $\hat{\bm{s}}$. The small group of $\rm K$ ($\Gamma$) is $C_3$ ($C_{3v}$).
Up to terms linear in momentum, which is here measured from K and $\Gamma$,
the SOC Hamiltonians compatible with those symmetries are
\begin{align}
{\cal H}_{\rm eff}^{\tau\rm K}&=\lambda_{\rm BR}^{\rm K}(k_x \hat{s}_y - k_y \hat{s}_x) +\tau\lambda_{\rm I}\hat{s}_z\,\label{Eq: eff_2x2_SOC},\\
{\cal H}_{\rm eff}^{\Gamma}&=\lambda_{\rm BR}^\Gamma(k_y \hat{s}_x - k_x \hat{s}_y)\,.
\end{align}
Here $\tau = 1$ ($-1$) stands for $\rm K$ ($\rm K'$), $\lambda_{\rm I}$ is the adatom-modified
intrinsic spin-orbit coupling and $\lambda_{\rm BR}$ is the adatom-induced
(Bychkov-Rashba-like as in semiconductor physics \cite{Bychkov1984:JETPL}) spin-orbit coupling. We will see that the latter comes from the space and pseudospin
inversion asymmetry. Contrary to graphene, the BR SOC at $\rm{K}$ depends on the momentum magnitude.
Higher-order terms in ${\cal H}_{\rm eff}^{\tau\rm K}$ and ${\cal H}_{\rm eff}^{\Gamma}$ are
presented in \cite{SM}.

Figure ~\ref{Fig:SHG} (bottom) shows the extracted
SOC parameters as functions of $\Delta/a$.
The intrinsic SOC $\lambda_{\rm I}$
is obtained from the splitting of the band at $\rm K$, see inset
in Fig.~\ref{Fig:SHG}(c). Parameter $\lambda_{\rm BR}^{\rm K}$
is extracted by fitting the linear dependence of the ratio of
the spin expectation values, $\langle\hat{s}_x\rangle/\langle\hat{s}_z\bigl\rangle|_{k_x=0} =
\lambda_{\rm BR}^{\rm K} k_y /\lambda_{\rm I} + O(k_y^3)$,
close to $\rm K$. The trigonally warped spin texture around K
is shown in the inset of Fig.~\ref{Fig:SHG}(d) and can be described
by higher-order terms in ${\cal H}_{\rm eff}^{\tau\rm K}$ \cite{SM}.
Finally, $\lambda_{\rm BR}^\Gamma$ is obtained
by fitting the spin splitting at $\Gamma$, see inset
in Fig.~\ref{Fig:SHG}(e). SOC is significantly enhanced in comparison 
to graphene. Directly comparable is the intrinsic SOC parameter whose
value in graphene is $2\lambda_{\rm I}=24\,\mu{\rm eV}$  \cite{Gmitra2009:PRB}.

The above single-band model can be obtained from a tight-binding (TB)
Hamiltonian using the carbon $p_z$ and hydrogen $s$ orbital basis.
The Hamiltonian contains orbital and SOC parts, $\mathcal{H}=\mathcal{H}_{\rm orb}+\mathcal{H}_{\rm so}$.
We denote by $c^\dagger_{i\sigma}=(a^\dagger_{i\sigma},
b^\dagger_{i\sigma})$ and $c_{i\sigma}=(a_{i\sigma}, b_{i\sigma})$
the creation and annihilation operators for the $p_z$ orbitals on the sublattices
($\rm A, \rm B$), with spin $\sigma$ and lattice site $i$.
Similarly, we define $h^\dagger_{m\sigma}$ and $h_{m\sigma}$ for
the hydrogen $s$ orbitals on adatom sites $m$. For the orbital part $\mathcal{H}_{\rm orb}$ we
take the TB model Hamiltonian introduced in Refs. \onlinecite{Robinson2008:PRL,Wehling2009:PRB},
which assumes the nearest neighbor carbon-carbon (C-C) hopping
$t=2.6$~eV, direct carbon-hydrogen (C-H) hopping $T$, and the adatom on-site energy $\varepsilon_{\rm h}$:
\begin{equation}\label{Eq:TB-SHG}
\begin{aligned}
\mathcal{H}_{\rm orb}=\varepsilon_{\rm h}\sum\limits_{m} \,h_{m\sigma}^\dagger\, h_{m\sigma}^{\phantom\dagger}+
T\sum\limits_{\langle m,i \rangle} \,h_{m\sigma}^\dagger\,c_{i\sigma}^{\phantom\dagger}
-t \sum\limits_{\langle i,j\rangle} \,c^\dagger_{i\sigma}\,c_{j\sigma}^{\phantom\dagger}.
\end{aligned}
\end{equation}
The angle brackets denote the nearest neighbors. Fitting the TB model to the
first-principles band structure for $\Delta/a=14\%$
(distortion in the single-adatom limit, see below) we obtain
$\varepsilon_{\rm h}=3$~eV and $T=6.5$~eV. The values are reliable in the vicinity 
of $\rm{K}$ point where $p_z$ carbon orbitals dominate the projected local DOS.

The SOC Hamiltonian can be derived by inspecting the reduction of the
graphene point group symmetry $D_{6h}$---which allows for the intrinsic spin-orbit coupling
$\lambda_{\rm{I}}$ only---to the one corresponding to SHG.
First, the C-H covalent bonds break the space inversion symmetry and the point group reduces to $C_{6v}$.
This structure inversion asymmetry induces the Bychkov-Rashba-like term
$\Lambda_{\rm BR}$. Second, the hydrogenated carbons
on sublattice A cannot be interchanged with the non-hydrogenated carbons
on sublattice B. This breaks the pseudospin inversion symmetry and $C_{6v} \rightarrow C_{3v}$.
The effect of the latter reduction is twofold:
(i)~The intrinsic SOC depends on the sublattice: $\Lambda_{\rm I}^{\rm A}$ and $ \Lambda_{\rm I}^{\rm B}$;
(ii)~New SOC terms emerge due to the pseudospin inversion asymmetry,
 $\Lambda_{\rm PIA}^{\rm A}$ and $\Lambda_{\rm PIA}^{\rm B}$, discussed below.

As the hydrogen $s$ orbitals do not directly contribute to SOC, we can express
the SOC TB Hamiltonian in the $p_z$ basis. In the next-nearest-neighbor limit
this Hamiltonian has five real parameters and reads,
\begin{equation}\label{Eq:SOC-SHG}
\begin{aligned}
&\mathcal{H}_{\rm so}=\frac{2i}{3}\sum\limits_{\langle i,j\rangle} c^\dagger_{i\sigma\phantom{'}} c_{j\sigma'}^{\phantom{\dagger}}\,
\bigl[\Lambda_{\rm{BR}}\bigl( \hat{\boldsymbol{s}}\times\boldsymbol{\mathrm{d}}_{\,ij}\bigr)_z\bigr]_{\sigma\sigma'}\\
&+\frac{i}{3}\sum\limits_{\llangle i,j\rrangle} c^\dagger_{i\sigma\phantom{'}} c_{j\sigma'}^{\phantom\dagger}
\Bigl[
\frac{\Lambda_{\rm{I}}^c}{\sqrt{3}}\,\nu_{ij}^{\phantom\dagger}\,\hat{s}_z +
2\Lambda_{\rm{PIA}}^c\bigl(\hat{\boldsymbol{s}}\times\boldsymbol{\mathrm{D}}_{ij}\bigr)_z
\Bigr]_{\sigma\sigma'}.
\end{aligned}
\end{equation}
The double angle bracket stands for the next nearest neighbors and
label $c$ denotes sublattice A or B. Factors $\nu_{ij}=1$ ($-1$) for
clockwise (counterclockwise) hopping path $j$ to $i$.
The nearest-neighbor $\boldsymbol{\mathrm{d}}_{ij}$ and next-nearest-neighbor
$\boldsymbol{\mathrm{D}}_{ij}$ unit vectors point from $j$ to $i$ (in a flat lattice).
The first term in Eq.~(\ref{Eq:SOC-SHG}) is the standard Bychkov-Rashba hopping as for graphene.
The second term describes the sublattice resolved intrinsic SOC
which couples same spins, and the PIA term
which couples opposite spins on the same sublattice.
Hamiltonian $\mathcal{H}_{\rm so}$ in Eq.~(\ref{Eq:SOC-SHG}) applies
to any hexagonal lattice system with $C_{3v}$ point group symmetry,
such as BN, or silicene in a transverse electric field.

The single-band limit described by Eq.~(\ref{Eq: eff_2x2_SOC}) can
be obtained from the TB Hamiltonian
$\mathcal{H}=\mathcal{H}_{\rm orb}+\mathcal{H}_{\rm so}$
by downfolding to sublattice B. This gives,
\begin{align}
\lambda_{\rm BR}^{\rm K}&\simeq-a\Lambda_{\rm PIA}^{\rm B}-\sqrt{3}\,a\Lambda_{\rm BR}\,\frac{t\,\varepsilon_{\rm h}}{T^2}\,,\\
\lambda_{\rm I}&\simeq-\Lambda_{\rm I}^{\rm B}-2\frac{\Lambda_{\rm BR}^2\,\varepsilon_{\rm h}}{T^2}\,.
\end{align}
Both PIA and BR SOC hopping terms contribute to the effective band SOC parameters.
This is the likely reason for the extracted nonmonotonic dependence of $\lambda_{\rm{I}}$ and the decrease
of $\lambda^{\rm{K}}_{\rm{BR}}$ as a function of $\Delta/a$ shown Fig.~\ref{Fig:SHG} (bottom),
The TB model cannot be reliably used at $\Gamma$ as there other bands (orbitals) mix in, see Fig. \ref{Fig:SHG}(a)
and Ref. \onlinecite{SM}.

We also present an effective SOC Hamiltonian close to $\rm{K}$. After
transforming $\mathcal{H}_{\rm so}$ to the ordered Bloch basis 
$[\psi_{\rm{A}\uparrow}(k),\psi_{\rm{A}\downarrow}(k),\psi_{\rm{B}\uparrow}(k),\psi_{\rm{B}\downarrow}(k)]$
and linearizing near $\rm{K}(\rm{K}')$ we obtain,
\begin{equation}\label{Eq:full-SHG-linearized}
\begin{aligned}
\mathcal{H}^{\tau\rm{K}}_{\rm so}=
&\,\Lambda_{\rm{BR}}\bigl(\tau\hat{\sigma}_x\hat{s}_y-\hat{\sigma}_y\hat{s}_x\bigr)\\
+&\,\tfrac{1}{2}\bigl[\Lambda_{\rm I}^{\rm A+B}\,\hat{\sigma}_z+\Lambda_{\rm I}^{\rm A-B}\,\hat{\sigma}_0\bigr]\tau\hat{s}_z\\
+&\,\tfrac{1}{2}\bigl[\Lambda_{\rm PIA}^{\rm A+B}\,\hat{\sigma}_z+\Lambda_{\rm PIA}^{\rm A-B}\,\hat{\sigma}_0\bigr]a(k_x\hat{s}_y-k_y\hat{s}_x)\,.
\end{aligned}
\end{equation}
Here $(\hat{\sigma}_0,\hat{\boldsymbol{\sigma}})$ and $(\hat{s}_0,\hat{\boldsymbol{s}})$
stand for the unit and Pauli matrices in the pseudospin and spin spaces, respectively.
The momentum is measured form $\rm K(K')$ and parameters
$\Lambda^{\rm A\pm B}=\Lambda^{\rm A}\pm\Lambda^{\rm B}$.
If $z$-inversion symmetry is restored, $\Lambda_{\rm BR}$,
$\Lambda_{\rm PIA}^{\rm A+B}$, and $\Lambda_{\rm I}^{\rm A-B}$ vanish
and one obtains the silicene limit \cite{Cheng-Cheng2011:PRB}.

\paragraph{Hydrogen on a supercell: single-adatom limit.}

The single-adatom limit is represented by a 5 $\times$ 5
supercell with a single hydrogen (2\% coverage). We use a fully relaxed structure with
$\Delta \approx 0.36$ $\rm \AA$ (14\% distortion) and the next nearest neighbor
distance $a$ = 2.516 $\rm \AA$ of the carbon
atoms around the hydrogenated site $\rm C_H$.

\begin{figure*}
\centering
\includegraphics[width=1.99\columnwidth,angle=0]{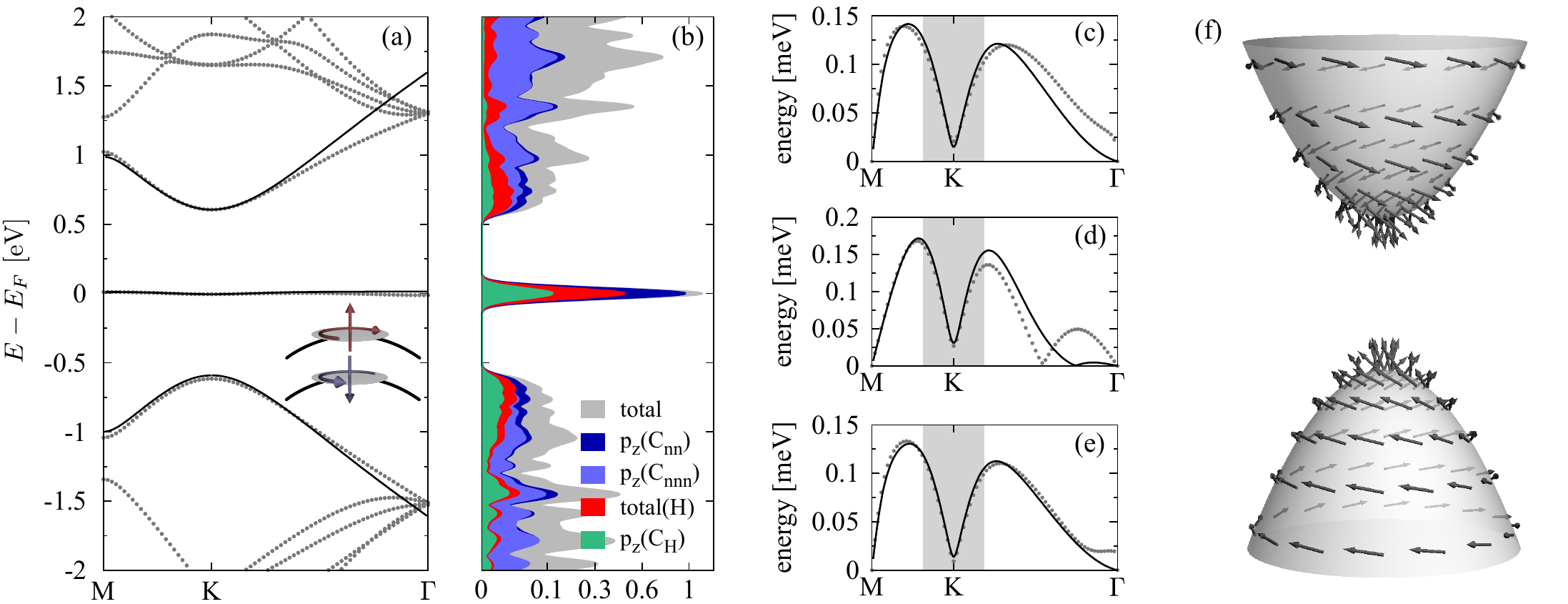}
\caption{(Color online) First-principles results and their tight-binding model fits
for hydrogen adatom on $5\times 5$ supercell.
(a)~Electronic band structure around the Fermi level. Dots are first-principles
results and solid lines are tight-binding model fits. The band
spin-orbit splitting around $\rm K$ is sketched for the valence band,
indicating the out of plane $z$ components and the rotation directions of
the in-plane spin components.
(b)~Broadened total density of states per atom (gray) and $p_z$ projected
local densities for atoms in the vicinity of the adatom: hydrogenated
carbon atom ${\rm C_H}$ (green), its nearest neighbor ${\rm C_{nn}}$
(dark blue), and the next-nearest neighbor ${\rm C_{nnn}}$ (light blue),
and the total density on hydrogen $\rm{H}$ (red). The projected densities of states
are normalized to the corresponding number of atoms in the set.
Conduction (c), impurity (d), and valence (e) band spin-orbit splittings
along the high-symmetry lines; symbols as in (a).
Tight-binding model least-square fits are performed within the shaded
regions around $\rm{K}$.
(f)~Spin expectation values around $\rm K$ for the spin-orbit split valence 
and conduction bands closer to the Fermi level. 
}
\label{Fig:bands_AB+TB}
\end{figure*}

Figure~\ref{Fig:bands_AB+TB}(a) shows the calculated spin-unpolarized
electronic band structure of our $5\times 5$ supercell.
The low energy spectrum contains three characteristic bands:
the valence and conduction bands, and the mid-gap impurity-like band.
These three bands can be nicely fitted by two parameters $T=7.5$~eV, and
$\varepsilon_{\rm h}=0.16$~eV entering the Hamiltonian
$\mathcal{H}_{\rm orb}$, Eq.~(\ref{Eq:TB-SHG}), as seen in
Fig.~\ref{Fig:bands_AB+TB}(a). Our values
differ from Refs. \onlinecite{Robinson2008:PRL, Wehling2009:PRB};
the comparison is discussed in \cite{SM}.
Larger supercells are also well described
by these parameters, confirming that the $5\times 5$ one
already describes the dilute limit, see \cite{SM}. The TB model with $s$ and $p_z$ orbitals
is also supported by the projected $p_z$ local density of states per carbon atoms
around the hydrogenated carbon as compared to the total density of states,
see Fig.~\ref{Fig:bands_AB+TB}(b).

\begin{figure}
\centering
\includegraphics[width=0.99\columnwidth,angle=0]{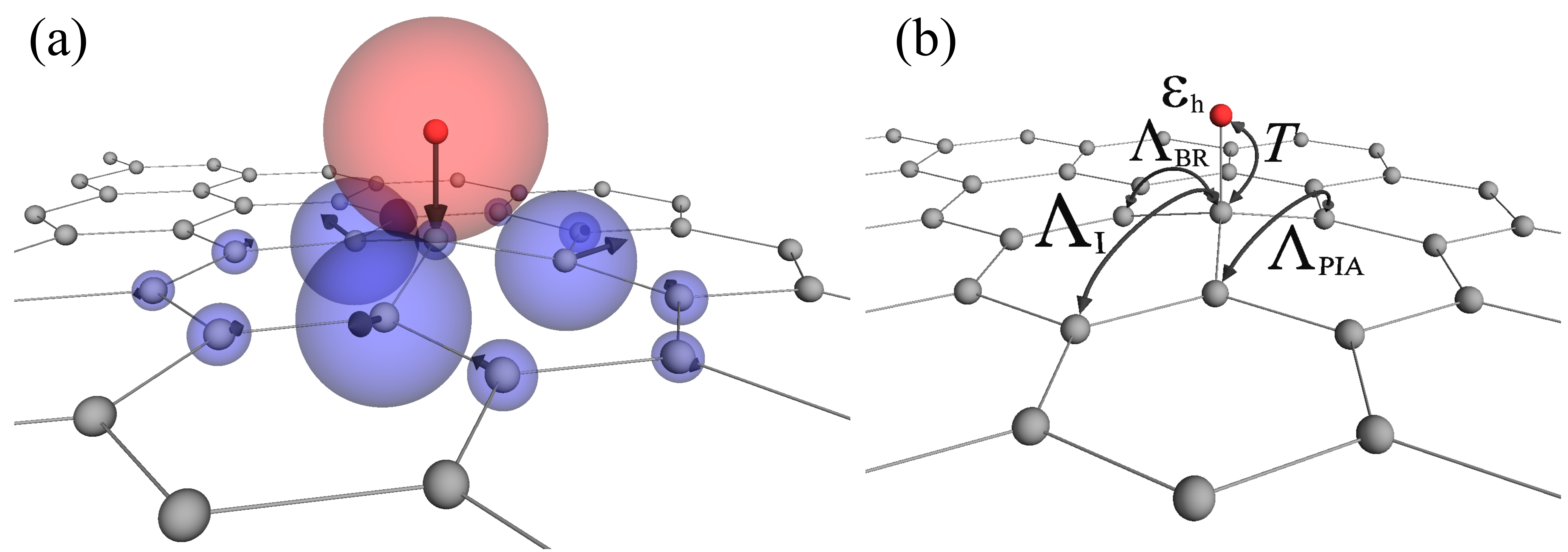}
\caption{(Color online)
(a)~First-principles calculations of electric dipole moments induced
by hydrogen adatoms on a $5\times 5$ supercell.
Directions of the dipole moments are shown by arrows; the sphere radii
correspond to the dipole magnitudes.
(b)~Hopping scheme of the tight-binding model showing the relevant
orbital and spin-orbit coupling parameters.
}
\label{Fig:dipoles+TB}
\end{figure}

The calculated spin-orbit splittings are shown in Fig. \ref{Fig:bands_AB+TB}(c).
To explain them we propose a minimal realistic SOC model which is locally $C_{3v}$ 
invariant in the impurity region with the C-H bond as the threefold axis of symmetry.
We deduce the impurity region from the induced dipole moments,
shown in Fig.~\ref{Fig:dipoles+TB}(a). The main effects are confined
up to the second nearest neighbors of the hydrogenated site
$\rm C_H$ (in sublattice A), defining our impurity region.
We use $A_\sigma^\dagger$ ($A_\sigma^{\phantom{\dagger}}$)
for the creation (annihilation) operators on $\rm C_H$ and
$B_{m,\sigma}^\dagger$ ($B^{\phantom{\dagger}}_{m,\sigma}$)
on the three nearest neighbors. Otherwise the terminology follows
the SHG case. The SOC Hamiltonian compatible with the local
symmetry is
\begin{equation}\label{Eq:SI-model_SOC}
\begin{aligned}
&\mathcal{H}_{\rm{so}}=\frac{i}{3}\sum\limits_{\llangle i,j\rrangle}\hspace{-1mm}{}^{{'}} c^\dagger_{i\sigma\phantom{'}} c^{\phantom{\dagger}}_{j\sigma'}
\Bigl[\frac{\lambda_{\rm{I}}}{\sqrt{3}}\,\nu_{ij}^{\phantom\dagger}\,\hat{s}_z\Bigr]_{\sigma\sigma'}\\
&+\frac{i}{3}\sum\limits_{\llangle {\rm C_H},j\rrangle} A^\dagger_{\sigma\phantom{'}} c_{j\sigma'}^{\phantom\dagger}
\Bigl[\frac{\Lambda_{\rm{I}}}{\sqrt{3}}\,\nu_{{\rm C_H},j}^{\phantom\dagger}\,\hat{s}_z\Bigr]_{\sigma\sigma'}+\mathrm{h.c.}\\
&+\frac{2i}{3}\sum\limits_{\langle {\rm C_H},j\rangle} A^\dagger_{\sigma\phantom{'}} B_{j\sigma'}^{\phantom{\dagger}}\,
\bigl[\Lambda_{\rm{BR}}\bigl( \hat{\boldsymbol{s}}\times\boldsymbol{\mathrm{d}}_{\,{\rm C_H},j}\bigr)_z\bigr]_{\sigma\sigma'}+\mathrm{h.c.}\\
&+\frac{2i}{3}\sum\limits_{\llangle i,j\rrangle}
B^\dagger_{i\sigma\phantom{'}}\,B_{j\sigma'}^{\phantom{\dagger}}
\bigl[\Lambda_{\rm{PIA}}^{\rm B}
\bigl(\hat{\boldsymbol{s}}\times\boldsymbol{\mathrm{D}}_{ij}\bigr)_z\bigr]_{\sigma\sigma'}.
\end{aligned}
\end{equation}
The first term is the graphene intrinsic SOC ($2\lambda_{\rm I}=24\,\rm{\mu eV}$).
It couples all next-nearest neighbors pairs not containing (this is denoted by the primed
summation symbol) $\rm C_H$. The second term describes the adatom induced
intrinsic spin-orbit coupling coupling $\Lambda_{\rm{I}}$, which couples the same spins
on the same sublattice. The third term, with Bychkov-Rashba hopping parameter
$ \Lambda_{\rm{BR}}$, describes the induced nearest neighbor spin flips. Finally,
the fourth term, with PIA parameter $\Lambda_{\rm{PIA}}$, comes from the pseudospin
inversion asymmetry. This term couples opposite spins of the next nearest
neighbors. We remark that $C_{3v}$ symmetry allows more spin-orbit hopping terms in
our impurity region. We considered them all but found only the three $\Lambda$'s 
in Eq.~(\ref{Eq:SI-model_SOC}) relevant to explain our \emph{ab-initio} results,
see the scheme in Fig.~\ref{Fig:dipoles+TB}(b).

Figures~\ref{Fig:bands_AB+TB}(c,d,e) show spin-orbit coupling induced
band splittings along high symmetry lines. The multiband least-square fit
around K point gives the following values for the
SOC parameters: $\Lambda_{\rm I}= -0.21$~${\rm meV}$, which is about
17 times larger than that of graphene $\lambda_{\rm I}$;
$\Lambda_{\rm BR} = 0.33$~meV, more than 60 times the value in graphene
 where $\lambda_{\rm BR} = 5$ $\mu$eV in a representative transverse electric field of 1 V/nm \cite{Gmitra2009:PRB};
$\Lambda_{\rm PIA}^{\rm B} = -0.77$~meV, which has no counterpart
in flat graphene.
The signs of the above parameters have been determined from the
spin expectation values around the $\rm K$ point, shown in Fig.~\ref{Fig:bands_AB+TB}(f).
The spin texture is governed mainly by PIA SOC.
Those parameters also fit larger supercells \cite{SM} and we propose
them, together with Hamiltonian Eq.~(\ref{Eq:SI-model_SOC}),
to describe the single adatom limit important for investigating
spin-flip and spin-orbit scattering in graphene.

In conclusion, we investigated spin-orbit coupling induced by hydrogen,
representing light adatoms, on graphene in dense and dilute limits.
We introduced realistic model spin-orbit Hamiltonians and provided
quantitative values for their parameters that can be used to study
spin relaxation, spin transport, and mesoscopic transport in
graphene with adatoms or in similar two-dimensional structures of the same symmetry.

We thank T.~O.~Wehling for helpful discussions. This work was supported by the 
DFG SFB 689, SPP 1285, and GRK 1570.

\bibliography{grp_adatom}


\clearpage

\pagestyle{empty}

\begin{widetext}

\includegraphics[width=1.0\columnwidth,angle=0]{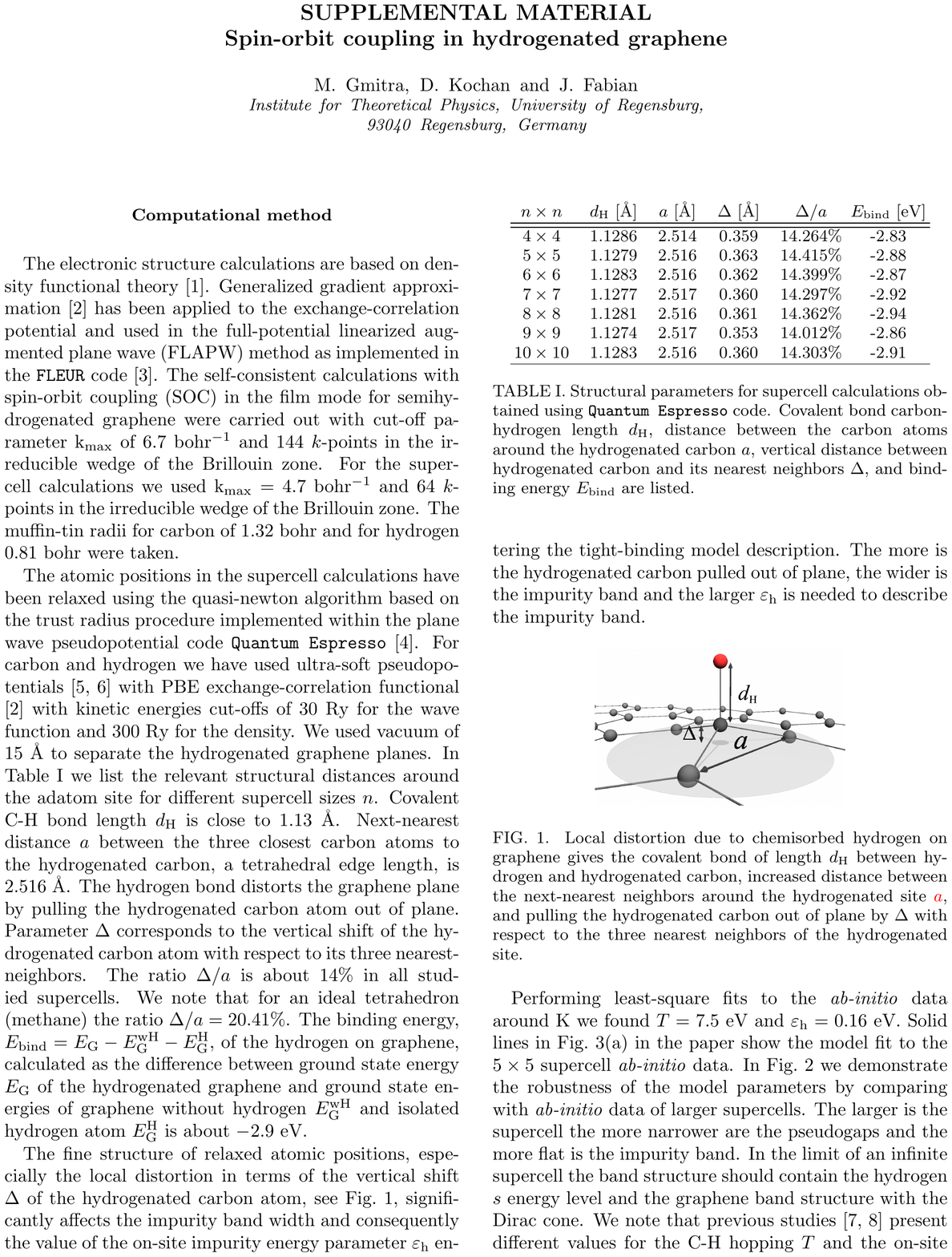}
\includegraphics[width=1.0\columnwidth,angle=0]{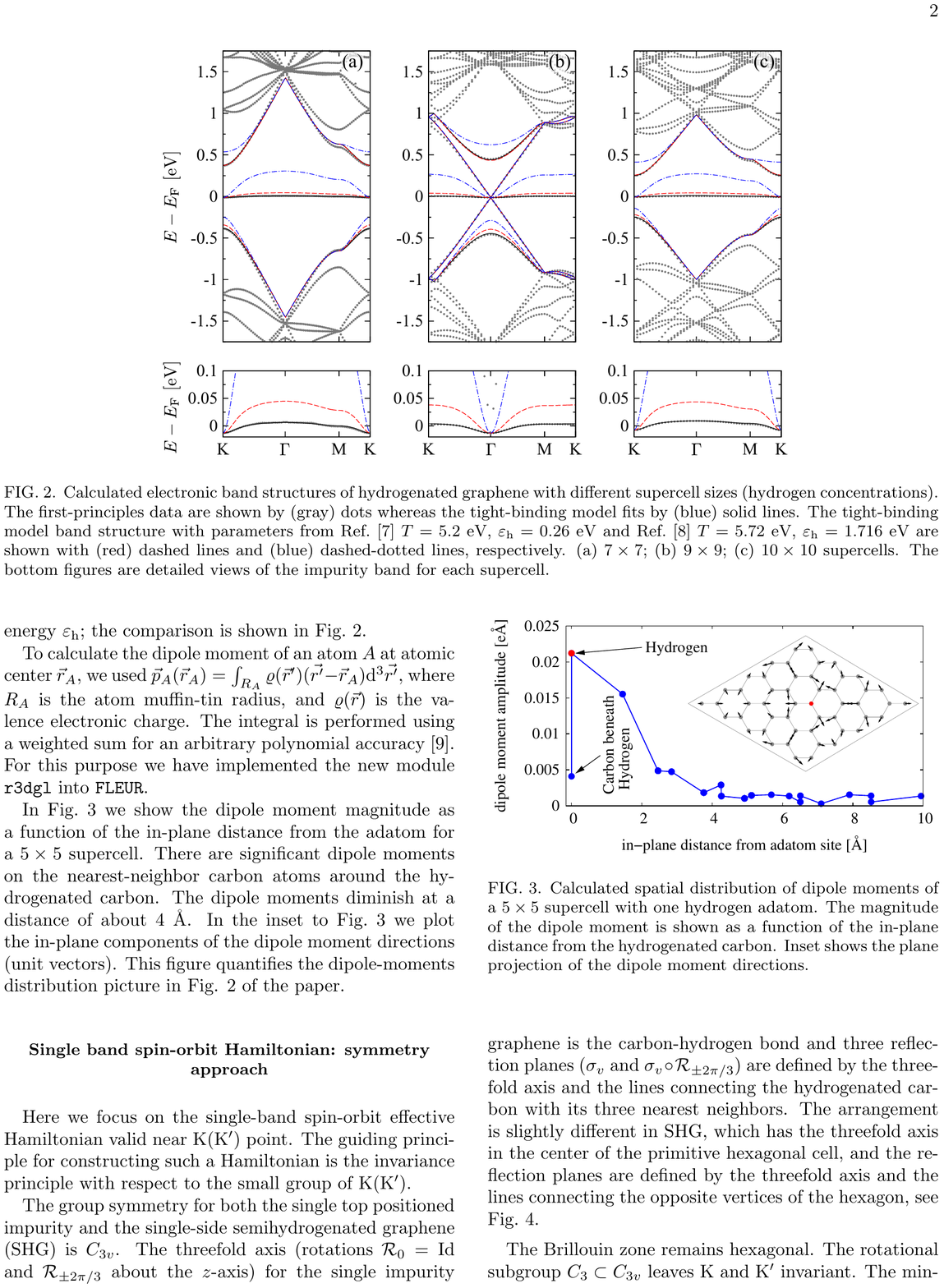}
\includegraphics[width=1.0\columnwidth,angle=0]{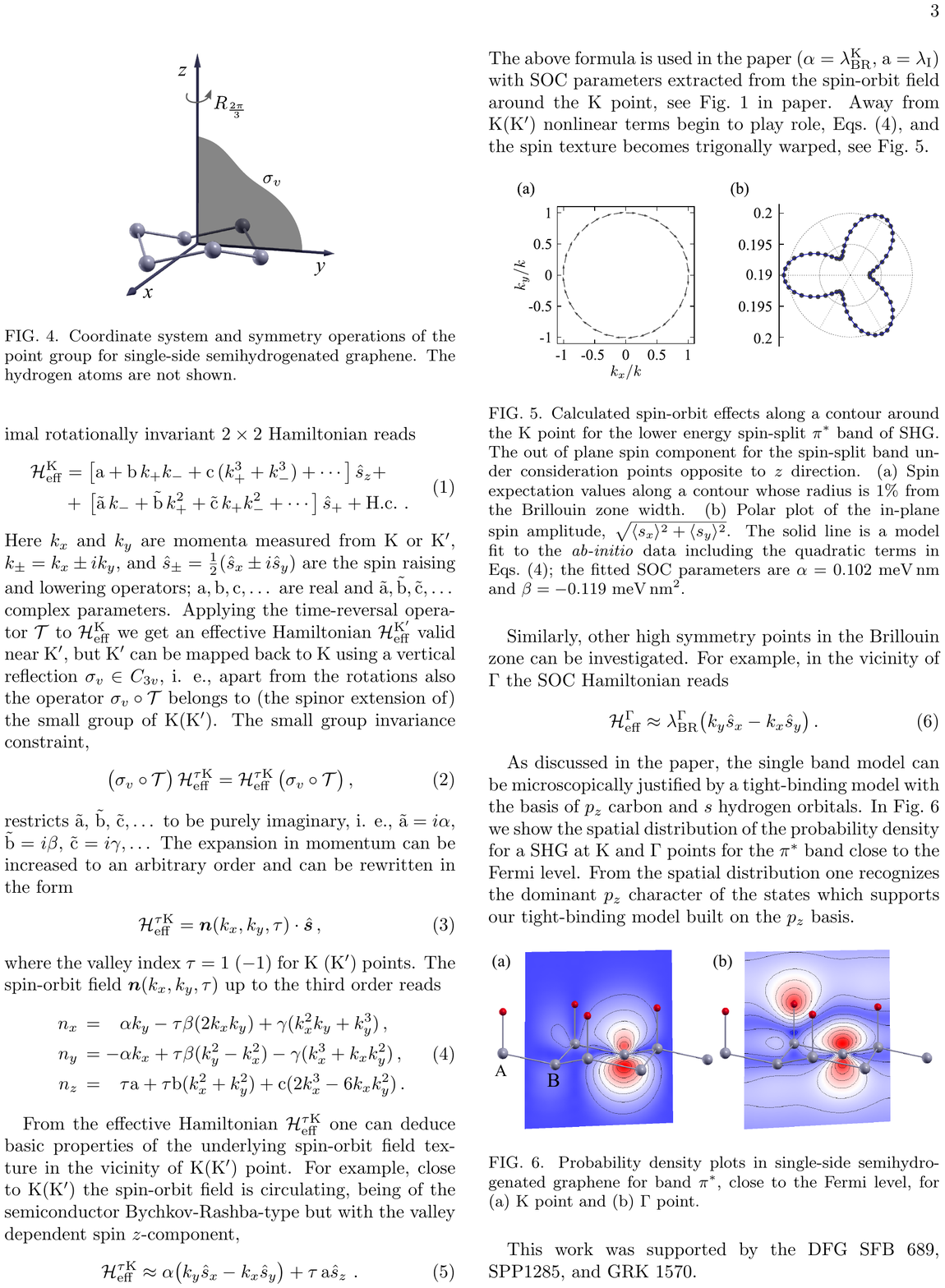}
\includegraphics[width=1.0\columnwidth,angle=0]{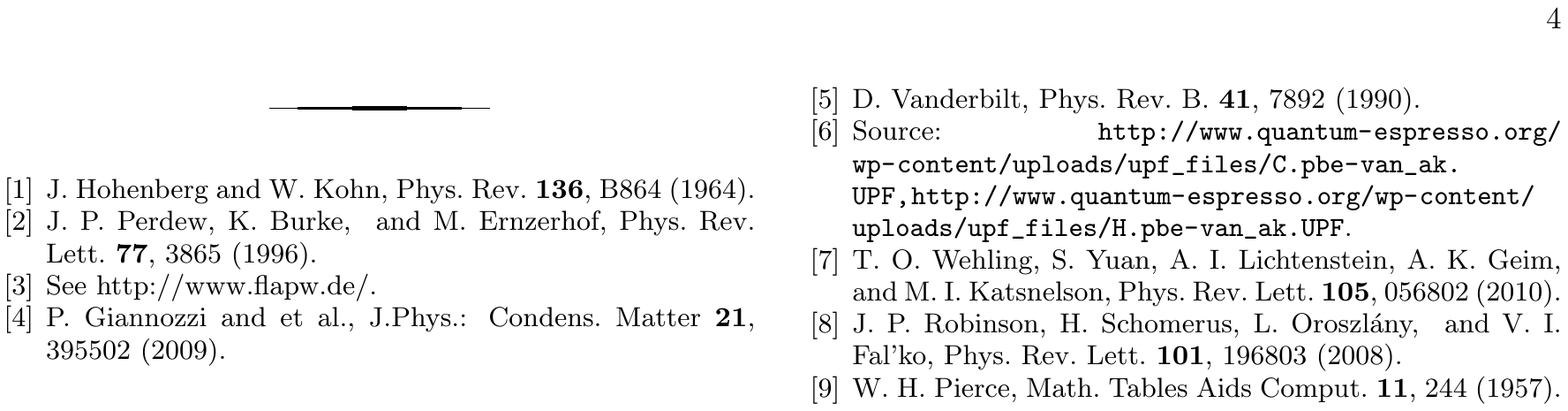}

\end{widetext}


\end{document}